\documentclass[12pt]{article}  

\let\ni=\noindent

\usepackage{graphicx}
\usepackage{natbib}
\usepackage{txfonts}
\newcommand{\cgs}{{\it cgs}}
\newcommand{\fdg}{\ensuremath{.\hspace{-3pt}^\circ}}
\begin{document}
\centerline{\Large{\bf The redshift distribution of the X-ray background}}

\vspace{2mm}
   \centerline {\large Andrzej M. So\l tan}

   \centerline{Nicolaus Copernicus Astronomical Center}

   \centerline{{\small Bartycka 18, 00-716 Warsaw, Poland}}
   \centerline{\small{soltan@camk.edu.pl}}

  \abstract
   \ni The X-ray background (XRB) is produced by a large number of faint sources
    distributed over a wide range of redshifts. The XRB carries information
    on the spatial distribution and evolution of these sources.
   {The goals of the paper are: 1. to determine the redshift distribution of the soft X-ray
    background photons produced by all types of extragalactic sources, in order
    to relate fluctuations of the background to the large scale structures,
    2. to determine the  redshift distribution of the soft XRB produced by AGN in
    order to calculate the evolution of the AGN X-ray luminosity density.}
   {A set of major X-ray surveys is used to determine the redshift
    distributions of the X-ray sources selected at various flux
    levels. Simple analytic fits to the data allow us to determine the
    smooth relationship between the redshift distribution and the
    source flux. The redshift distribution of the integral XRB
    flux is obtained  by averaging the fits over the source counts.}
   {It is shown that the distribution of extragalactic XRB photons in the
    $0.5-2$\,keV band is adequately represented by the function:
    $d\,n_{\rm XRB} / d\log z = 5.24\:z^{1.52}\,\exp(-z/0.63)$.
    The huge voids postulated to explain the cold spots
    in the CMB maps create dips in the total XRB flux.
    However, the expected magnitude of the effect is comparable to the
    fluctuation amplitude of the XRB generated by the individual sources
    contributing to the background.
    The cosmic evolution of the AGN X-ray luminosity density up to
    redshift of $\sim\!5$ is calculated in an elegant and straightforward
    way. Systematic uncertainties of the present method
    are assessed and shown to be small.
    At redshift greater than one the present results could be compared
    directly with some recent estimates obtained in a standard way and
    the agreement between both methods is very good.}

\vspace{2mm}
\ni   Keywords{X-rays: diffuse background  --
             intergalactic medium   --
             X-rays: galaxies }

\section{\large Introduction \label{intro}}

The X-ray background (XRB) is generated mostly by discrete extragalactic sources
(e.g. \citealt{lehmann01, kim07}, and references therein), predominantly by
various types of Active Galactic Nuclei (AGN) and cluster of galaxies. A
question of flux, luminosity and redshift distributions of these sources has
been discussed in a great number of papers for the last $30$ years. One of the
major outcome of these investigations is the conclusion that X-ray sources
associated with the AGN are subject to strong cosmic evolution (e.g.
\citealt{miyaji00}, \citealt{silverman07}, and references therein).
As a result of the evolution, the redshift distribution of the XRB flux
is wide. Thus, the integral XRB comprises the information on the
large scale distribution of the X-ray sources over a wide redshift range.

In the present paper the redshift distribution of the XRB photons is
investigated in detail. The analysis is based on an extensive observational
data selected from several published X-ray sky surveys. A convenient analytic
approximations are applied to model the observed redshift histograms of the
extragalactic X-ray sources selected at several flux levels. These
distributions are weighted by the source counts and summed up to obtain the
redshift distributions of the integral XRB. 

Next, the redshift distribution is used to define a relationship between the
XRB signal and the large-scale fluctuations of the matter spatial distribution.
The investigation has been raised by the recent report on the huge void
generating a dip in the surface brightness of the radio background
(\citealt{rudnick07}). This void, responsible for the deficit of the radio
surface brightness, allegedly generates also a cold spot in the CMB map via the
late-time integrated Sachs-Wolfe effect. Although the careful statistical
analysis by \cite{smith08} has not confirmed the existence of this
particular ``cold spot'' in the radio survey, a relationship between the large
scale features of the matter distribution and the integrated sky brightness in
various energy bands is a problem deserving some interest. 

Apart of the question of the XRB fluctuations induced by voids, the XRB
redshift distribution is interesting per se, as it allows to assess the
evolution of AGN phenomenon. A standard way to estimate a rate and type of this
evolution is based on the examination of the X-ray luminosity functions
determined at the consecutive redshift bins.  Unfortunately, the X-ray surveys
produce flux-limited rather than luminosity-limited samples of sources. In
effect, luminosity functions at different redshifts cover different luminosity
ranges. This in turn severely impedes estimates of the luminosity function over
a wide range of luminosities and redshifts.  The total level of nuclear
activity in galaxies within unit volume is given by the integral of the X-ray
luminosity function. The question of the AGN cosmic evolution constitutes one
of the central problems of observational cosmology, and has been investigated
for the last forty years (this issue was for the first time recognized by
\citealt{schmidt68}).  Here a question of the AGN evolution is addressed
without the  calculations of the X-ray luminosity function.  The available
observational data on X-ray source counts and redshifts are used to evaluate
the redshift distribution as a function of source flux. This relationship and
the source counts allow to calculate the redshift distribution of the total XRB
and the integral luminosity density generated by the AGN as a function of
redshift.

The organization of the paper is following. First, I present the formulae used
in calculations. Next, in Sec.\ref{catalogs}, the basic information on the
observational material extracted form the various archives is given. Since the
comprehensive characteristics of the data and the source catalogs are described
in the original papers, only the basic properties of the material are presented
here. The numerical fits to the observed distributions are obtained in
Sec.~\ref{fits}. In that section the calculations of the redshift distribution
of the XRB photons are described in details. These results are applied in the
Sec.~\ref{supervoids} to quantify the relationship between the voids and the
XRB variations. In Sec.~\ref{agn} the distribution of the XRB flux produced by
AGN is used to calculate the evolution of the AGN activity. Finally, potential
sources of errors inherent in the present method are discussed in the
Sec.~\ref{discussion}.

The `canonical' standard cosmology is assumed throughout, with $H_0 =
70$\,km\,s$^{-1}$Mpc$^{-1}$, $\Omega_m = 0.3$, and $\Omega_\Lambda = 0.70$. 

\section{\large Basic relationships \label{formulae}}

In the present approach the X-ray source catalogs are used to construct
redshift distributions of the extragalactic sources as a function of the source
flux. At this stage, a question of the source (absolute) luminosities is not
addressed. Let $N(S)$ denotes the X-ray source counts, i. e. number of sources
brighter than $S$ in a unit solid angle, and $f_S(z) = dn(z\!\mid\!S)/d\log z$
is the redshift distribution of sources with flux $S$.  Then, the redshift
distribution of the XRB surface brightness, $b(z)$, is equal to:

\begin{equation}
b(z) = \frac{1}{b} \int\! dS\,f_S(z)\:S \left|\frac{dN(S)}{dS}\right|\,,
\label{bz}
\end{equation}
where the integration covers the entire ``interesting'' range of source
fluxes $S$ and $b$ denotes the integral background flux:
\begin{equation}
b = \int\! b(z)\;d\log z = \int\! dS \,S \left|\frac{dN(S)}{dS}\right|\,.
\label{b}
\end{equation} 
It is assumed that the $f_S(z)$ distributions are normalized:
\begin{equation}
\int\! f_S(z)\;d\log z = 1\,.
\end{equation}
Here the integration limits cover the total range of redshifts occupied
by X-ray sources. The actual limits of the ``interesting'' range of
fluxes is discussed below.

The luminosity density, $\varepsilon(z)$, i. e. a total luminosity $L$ 
generated in a unit comoving volume, $V$:

\begin{equation}
\varepsilon(z) = \frac{dL}{dV}\,,
\end{equation}
is related to the flux distribution $b(z)$ and the luminosity distance,
$D_L(z)$:

\begin{equation}
\varepsilon(z)\;\frac{dV}{d\log z} = 4\,\pi\,D_L^2(z)\:b(z)\,.
\label{epsilon}
\end{equation}

The cosmological relationships between the comoving volume and the luminosity
distance in a flat space with $\Lambda \neq 0$ is given by \cite{hogg99}:

\begin{equation}
\frac{dV}{dz} = \frac{c}{H_0}\:\frac{D_L^2}{(1+z)^2
                \sqrt{\Omega_m (1+z)^3 + \Omega_\Lambda}}\,.
\label{dvdz}
\end{equation}
Combining Eqs.~\ref{epsilon} and \ref{dvdz} we finally get:
\begin{equation}
\varepsilon(z) = 4\pi\:\frac{H_0}{c}\;\frac{(1+z)^2
                 \sqrt{\Omega_m(1+z)^3 + \Omega_\Lambda}}{\ln\!10\;z}\; b(z)\,.
\label{lum_dens}
\end{equation}

Thus, to calculate the distributions $b(z)$ and $\varepsilon(z)$, the source
counts $N(S)$ and the functions $n(z\,|\,S)$ have to be determined using the
observational material. In the next section the available X-ray surveys are
examined from this point of view.

\section{\large Observational material \label{catalogs}}

Because the high imaging efficiency of X-ray telescopes in the soft band and
numerous extensive identifications programs, we concentrate on the X-ray band
of $0.5 - 2.0$\,keV. At these energies a fraction of the background resolved
into discrete sources exceeds $90$\,\% and is higher than in the other bands
(e.g. \citealt{moretti03}, \citealt{brandt05}).  Also a fraction of identified
objects with measured spectroscopic or photometric redshifts is relatively
high.

Equation~\ref{bz} shows that the $b(z)$ distribution is sensitive to sources
which perceptibly contribute to the XRB.  Consequently, one needs to calculate
the $n(z\,|\,S)$ functions over a quite wide range of fluxes. To achieve this
objective I have examined numerous X-ray surveys and selected  several major
source catalogs for further analysis. The overall characteristics of those
catalogs are listed in Table~\ref{surveys}. A common name of the survey/catalog
is given in column 1. From the each catalog, sources for further processing
have been extracted within fixed range of fluxes defined on column 2.  The
numbers of all sources, extragalactic and AGN with known redshifts are given in
columns 3, 4, and 5, respectively.

Statistical requirements which have to be satisfied by the source samples to
properly determine the $n(z\,|\,S)$ distribution are different than those for
the luminosity function calculations. The individual sample should contain
sources from possible narrow range of fluxes, but the sample has not to be flux
limited. The sample provides unbiased estimate of the redshift distribution as
long as the process of identification and redshift measurements does not
introduce spurious correlation between flux and redshift.

\begin{table}
\caption{The X-ray surveys selected for the analysis}
\label{surveys}
\centering
\begin{tabular}{llccc}
\hline\hline
\noalign{\smallskip}
 Name &\multicolumn{1}{c}{Flux limits}& \multicolumn{3}{c}{Number of sources} \\
      &\multicolumn{1}{c}{(erg\,cm$^{-2}$ s$^{-1})$}& All &Extragalactic& AGN \\
\hline
\noalign{\smallskip}
RBS   & $1.0\times 10^{-12}-5.0\times 10^{-11}$ & 1764 & 1054  & 681 \\
NEP   & $5.0\times 10^{-14}-1.0\times 10^{-12}$ &  361 &  248  & 192 \\
RIXOS & $2.5\times 10^{-14}-5.0\times 10^{-13}$ &  393 &  318  & 235 \\
XMS   & $1.0\times 10^{-14}-2.0\times 10^{-13}$ &  275 &  256  & 231 \\
CDFS  & $5.0\times 10^{-17}-1.0\times 10^{-15}$ &  205 &  201  & 197 \\
CDFN  & $1.5\times 10^{-17}-5.0\times 10^{-15}$ &  425 &  412  & 268 \\
\hline
\noalign{\smallskip}
\end{tabular}
\end{table}

\subsection{\normalsize The {\it ROSAT} Bright Survey (RBS)}

The identification program of the brightest sources detected in the {\it ROSAT}
All-Sky Survey, known as {\it ROSAT} Bright Survey, resulted in a sample of
2072 sources with the total count rate above $0.2$\,s$^{-1}$
(\citealt{schwope00}).  More than $99.5$\,\% of sources in the final catalogue
is identified.  The survey covers high galactic latitudes ($|b|> 30\deg$).
After the removal of the Virgo clusters and Magellanic Clouds regions, the
catalog contains 2012 sources.  In the energy band of $0.5-2.0$\,keV, 1773 RBS
sources generate flux between $1.0\times 10^{-12}$\,erg\,cm$^{-2}$s$^{-1}$
(hereafter \cgs) and $5.0\times 10^{-11}$ \cgs.  As one might expect, only for
a small fraction of the RBS sources the redshifts are undetermined and
relatively large number of sources is associated with galactic sources, mostly
late type stars and cataclysmic variables.

Since the RBS sample covers rather wide range of fluxes, it is useful for our
purposes to divide it into several subsamples with narrow flux limits and to
estimate the $n(z\,|\,S)$ distribution for the each set separately. We define
the bright source sample, labeled RBS(b), which contains sources with
$4.0\times 10^{-12} < S < 1.0\times 10^{-11}$ \cgs. Of $365$ RBS sources in
this flux range, $129$ is identified with galactic objects and for the other
$15$ sources redshifts have not been measured. Eventually, the sample contains
$221$ extragalactic objects with known redshifts. Nearly half of the sample,
viz. $97$ sources are identified with clusters of galaxies and normal galaxies.
The median flux\footnote{The samples extracted from the RBS span over narrow
ranges of fluxes and there is no major difference between the mean and median
values. In some other samples investigated in this paper the median flux is
distinctly smaller than the average. In those cases, the median flux is adopted
as the argument in the $n(z\,|\,S)$ functions.} in the extragalactic subsample
$S{\rm_m} = 5.5\times 10^{-12}$ \cgs.

\begin{figure*}
\begin{center}
\includegraphics[width=0.9\textwidth]{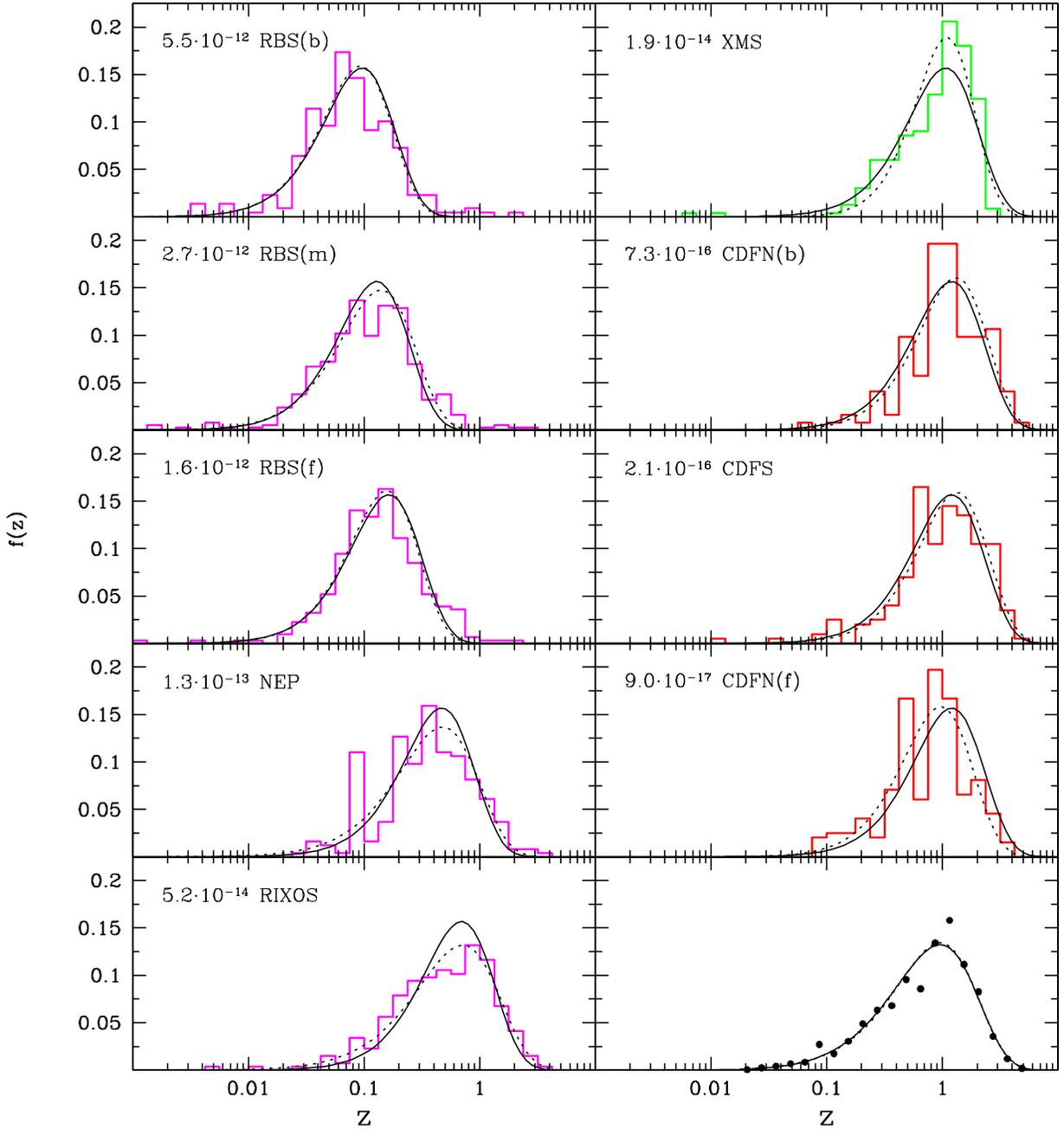}
\caption{\leftskip 10mm \rightskip 10mm
{\small Histograms with labels -- the redshift distributions of extragalactic
sources in $9$ samples constructed in the present investigation.  The data are
binned with $\Delta\log z = 0.125$; The integrals of all the histograms are
normalized to unity.  The sample designations and the median fluxes in \cgs\
are given in the upper left corners; dotted curves -- best 3 parameter
fits for the each sample separately; solid curves -- the best model for all
the samples. The lower right panel: solid curve -- the model redshift
distribution for the whole XRB (the normalization of the distribution is the
same as for the histograms); dotted curve -- analytic approximation of the
model distribution; points -- the distribution obtained by the linear
interpolation between the raw histograms (see text for details).}}
\label{z_distr}
\end{center}
\end{figure*}

The flux limits of $2.0\times 10^{-12} < S < 4.0\times 10^{-12}$ \cgs\ have
been adopted for the medium flux sample, RBS(m). Within this flux range the RBS
comprises of $669$ objects including $272$ galactic stars. Among the $397$
extragalactic sources, $233$ are identified with AGN with known redshifts. The
median flux in the RBS(m) extragalactic sample, $S_{\rm m} = 2.7\times
10^{-12}$ \cgs.  The faint source sample, RBS(f), contains sources with
$1.0\times 10^{-12} < S < 2.0\times 10^{-12}$ \cgs.  Of $599$ RBS sources in
this flux range $268$ is identified with galactic objects and $24$ has unknown
redshifts. Thus, the RBS(f) sample contains $307$ sources extragalactic
objects with the median flux of $1.6\times 10^{-12}$ \cgs.

The redshift distributions of sources in the RBS (b), (m), and (f) samples are
shown in three upper left panels in Fig.~\ref{z_distr}. The integrals of the
histograms for all the samples in Fig.~\ref{z_distr} are normalized to unity.
The distributions are plotted using logarithmic redshift bins with $\Delta \log
z = 0.125$. Each histogram is labeled with the survey name and the median flux
in \cgs.  Analytic fits will be discussed in the next section.

\subsection{\normalsize The {\it ROSAT} North Ecliptic Pole Survey (NEP)}

The deepest exposure of the {\it ROSAT} All-Sky Survey (RASS) is centered at
the north ecliptic pole (\citealt{voges99}). The RASS of this region has been
used to construct statistically well defined sample of X-ray sources above a
flux limit $\sim\!2\times 10^{-14}$ \cgs\ (\citealt{henry06}) which have been
followed-up by the optical observations (\citealt{gioia03}). The identification
rate in the final catalog of $443$ sources is very high ($99.6$\,\%).

Within the flux limits of $5.0\times 10^{-14}$ and $1.0\times 10^{-12}$ \cgs\
the NEP survey provided $361$ sources.  After excluding $113$ galactic stars, we
are left with $248$ extragalactic sources; for $3$ sources the redshift is
unknown. The redshift histogram of $245$ sources (including $53$ clusters) is
shown in Fig.~\ref{z_distr}; the median flux in this sample is equal to
$1.2\times 10^{-13}$ \cgs.

\subsection{\normalsize The {\it ROSAT} International X-ray/Optical Survey (RIXOS)}

This {\it ROSAT} medium-sensitivity survey consists of sources found
in $82$ PSPC pointing observations at high galactic latitudes ($|b|>28\deg$).
A flux limit of $3\times 10^{-14}$ \cgs\ was adopted in $64$ fields and
$8\times 10^{-14}$ \cgs\ in the remaining $18$ fields. The source selection
procedures, optical identifications and the final catalog are given by
\cite{mason00}.

For the purpose of the present analysis, $393$ sources with fluxes between
$2.5\times 10^{-14}$ and $5\times 10^{-13}$ \cgs\ have been selected. Within
these flux limits $75$ sources are associated with galactic stars. Of the
remaining $318$ sources, the redshifts of three objects are unknown, and $49$
sources are still unidentified. The redshift distribution of $266$ sources
(including $33$ clusters) is shown in Fig.~\ref{z_distr}. The median flux in this
subsample $S_{\rm m} = 5.2\times 10^{-14}$ \cgs.

\subsection{\normalsize The {\it XMM-Newton} serendipitous survey (XMS)}

The {\it XMM-Newton} serendipitous survey (XMS) has been constructed in a
similar way as the {\it RIXOS}. More than $300$ sources have been isolated in
$25$ high galactic latitude ($|b|>22\deg$) pointings covering
$\sim\!3$\,deg$^2$ of the sky (\citealt{barcons07}). In the $0.5-2.0$\,keV band
the sample is complete above $1.5\times 10^{-14}$ \cgs, and contains many
weaker sources.

I have extracted from the original catalog $275$ sources with fluxes between
$1.0\times 10^{-14}$ and $2.0\times 10^{-13}$ \cgs. The sample is completely
identified; it contains $19$ stars and $2$ clusters of galaxies. For $23$
objects the redshifts are unknown. The redshift distribution of $233$ sources
is shown in Fig.~\ref{z_distr}.  The median flux in this subsample $S_{\rm m} =
1.9\times 10^{-14}$ \cgs.

\subsection{\normalsize The Chandra Deep Field--South (CDFS)}

The  $1$\,Ms {\it Chandra} observations known as the {\it Chandra} Deep Field
South are described by \cite{giacconi02}. The catalog of sources detected in
this field by two independent algorithms contains $304$ objects, of which $275$
have determined fluxes in the $0.5-2.0$\,keV band. For further processing $205$
sources with fluxes in the range $5.0\times 10^{-17} - 1.0\times 10^{-15}$
\cgs\ have been selected. Four sources are identified with stars. The redshifts
either spectroscopic (\citealt{szokoly04}, \citealt{ravikumar07}) or
photometric (\citealt{zheng04}) are known for $200$ sources; one source remains
unidentified.

The redshift distribution is shown in Fig.~\ref{z_distr}. The median flux in
this sample $S_{\rm m} = 2.1\times 10^{-16}$ \cgs. The sample contains one
galaxy group; two other sources apparently are not associated with the activity
in the galactic nuclei (\citealt{lehmer06}). It is likely, however, that more
objects in the CDFS survey should be classified as off-nuclear sources.  The
available data do not allow for unambiguous separation of AGN and off-nuclear
sources at the low flux levels in the CDFS. This reservation holds also for the
CDFN samples below.

\subsection{\normalsize The Chandra Deep Field--North (CDFN)}

The ultra deep Chandra field, $2$\,Ms exposure, CDFN, resulted in a catalog
of $503$ sources  detected over $0.12$ sq.\,deg. (\citealt{alexander03}).
Optical follow-up observations by \cite{barger03} have rendered a large
number of spectroscopic and photometric redshifts. Several more redshifts
are taken from \cite{reddy06}, \cite{donley07} and \cite{georgakakis07}.

In the present investigation,
the CDFN catalog has been divided into two samples of bright (b) and faint (f)
sources. The (b) sample contains $181$ sources between $2.5\times 10^{-16}$ and
$5.0\times 10^{-15}$ \cgs; ten sources have been identified with the galactic
stars; at least one has been categorized as `starburst' galaxy
(\citealt{georgakakis07}), for $49$ objects the redshifts have not been
measured. The sample consists of $122$ sources, mostly AGN. The median flux
in this subsample $S_{\rm m} = 7.3\times 10^{-16}$ \cgs.

The faint CDFN sample has been selected between $1.5\times 10^{-17}$ and
$2.5\times 10^{-16}$ \cgs. Among $244$ sources satisfying these flux
limits, three sources have been identified with stars, $51$ -- with the
starburst galaxies (\citealt{georgakakis07}) and for $43$ objects the redshifts
have not been measured. The final sample used in the calculations contains $198$
sources with $147$ confirmed AGN. The median flux in the sample
$S_{\rm m} = 9.0\times 10^{-17}$ \cgs. The redshift
histograms for the (b) and (f) samples are shown in Fig.~\ref{z_distr}.

The numbers of objects unidentified or without redshift are in some samples
quite large. Hence one could expect that the corresponding redshift histograms
are not representative for the whole population of sources at given flux. Below
this question is de facto worked out where we construct an analytic function
which simultaneously fits all the histograms.

\section{\large Approximations and fits \label{fits}}

The redshift distribution of sources selected at fixed flux, $n(z|S)$,
is a intricate function of a number of parameters, such as the luminosity
function, the relationship between the luminosity and observable flux, and the
relationship between the volume and redshift. The luminosity function itself
depends on redshift and both the latter relationships depend on the cosmological
model. However, the existing estimators of the $n(z|S)$ function represented by
the nine histograms in Fig.~\ref{z_distr} are strongly affected/degraded by the
statistical noise. It implies that a simple analytic function with $2-3$ free
parameters will provide a statistically satisfactory fit to the observed
distributions.

It appears that the histograms in Fig.~\ref{z_distr} are adequately reproduced
by:
\begin{equation}
f_S(z) = f_0\;z^\alpha\:e^{-z/z_{\rm c}}\,,
\label{fit_1}
\end{equation}
where $f_0=f_0(S)$, $\alpha = \alpha (S)$, and $z_{\rm c}=z_{\rm c}(S)$ are
three parameters fitted to the the histograms $n(z|S_i)$, $i = 1,... ,9$. In
Fig.~\ref{z_distr} the least square fits for all the distributions are shown
with the dotted curves. Apart from a few pronounced features visible in the
plots which represent the large scale structures reported in the literature
(e.g.  \citealt{barger02}, \citealt{gilli03}), analytic fits seem to adequately
reproduce the observed distributions.

\begin{figure}
\begin{center}
\includegraphics[width=0.6\textwidth]{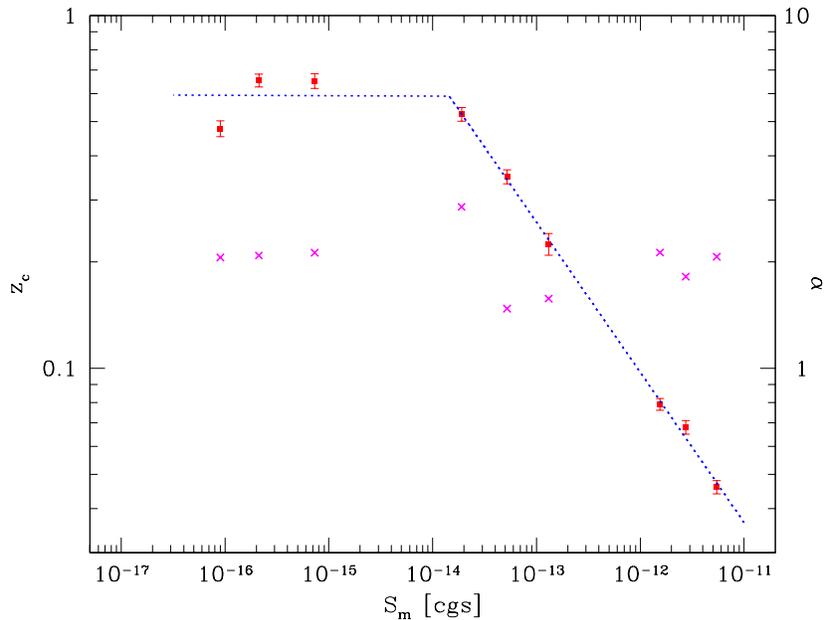}
\caption{\leftskip 10mm \rightskip 10mm
{\small Distributions of the best fitted parameters $\alpha$ and $z_{\rm c}$
vs. the median flux, $S_{\rm m}$. Points with error bars -- $z_{\rm c}$
(left-hand ordinate), crosses -- $\alpha$ (right-hand ordinate). The assumed
$z_{\rm c} \sim S$ relationships used to model the XRB redshift distribution is
shown with the dotted lines (see text for details).}}
\label{z_c_vs_s}
\end{center}
\end{figure}

It is found that only $z_{\rm c}$ is strongly correlated with $S_{\rm m}$,
while the fits do not indicate any statistically significant correlation
between $\alpha$ and $S_{\rm m}$. In Fig.~\ref{z_c_vs_s} the best fit values of
$\alpha$ are shown with crosses. The labels and scale on the left-hand ordinate
refer to $z_{\rm c}$, and on the right-hand -- to $\alpha$.  Since the
simultaneous fitting of $\alpha$ and $z_{\rm c}$ introduces a spurious
correlation between these two parameters, the $\alpha$ parameter has been fixed
at the average value found for the $9$ samples, $\bar{\alpha} = 1.934$.
Effectively, it means that the shape of the $n(z|S)$ function is fixed and the
only dependence on $S$ is limited to the horizontal shift along the $z$ axis.
In Fig.~\ref{z_c_vs_s} the best fit parameters $z_{\rm c}$ found for the
fixed $\alpha$ are plotted against the median flux $S_{\rm m}$.  In agreement
with the expectations, the $z_{\rm c}$ increases with diminishing flux $S_{\rm
m}$ over a wide range of fluxes.  However, a clear flattening of the
relationship is observed below $\sim\!10^{-14}$ \cgs.  This apparent absence of
correlation between $z_{\rm c}$ and $S$ results from the well-known fact that
in the X-ray surveys at low flux levels the maximum detected redshift remains
stable while significantly increases fraction of intrinsically weak sources.

The points in the $z_{\rm c} - S_{\rm m}$ relationship above $\sim 10^{-14}$
\cgs\ seem to follow the power law. In the subsequent calculations it is
assumed that this relationship is in fact well approximated by the power law in
the whole range of fluxes between $10^{-14}$ and $10^{-11}$ \cgs, although the
data coverage is rather sparse.  Below $S \approx 10^{-14}$ \cgs\ the data are
insufficient to delineate precisely the $z_{\rm c} - S_{\rm m}$ relationship. I
have assumed tentatively that $z_{\rm c}$ remains constant and is equal to the
average value found for three {\it Chandra} samples. The $z_{\rm c}$ in
CDFN(b), CDFS and CDFN(f) are equal to $0.69$, $0.69$ and $0.51$, respectively.
The average weighted by the uncertainties $\bar{z}_c = 0.61$. In
Fig.~\ref{z_c_vs_s} the model relationship $z_{\rm c} - S$ used in the
calculations is shown with the dotted line.

The $z_{\rm c} \sim S$ relationship with the fixed $\alpha$ parameter and
fixed normalization of the integral:
\begin{equation}
\int f_S(z)\;d\log z = 1\,,
\end{equation}
eliminates formally any free parameters in the fitting the analytic function
to the nine histograms in Fig.~\ref{z_c_vs_s}:
\begin{equation}
f_S(z)=\frac{\ln 10}{\Gamma(\alpha)\;z_{\rm c}^\alpha}\;z^\alpha\;e^{-z/z_{\rm c}}\,,
\label{fit_2}
\end{equation}

where $\alpha = 1.934$, $\Gamma(\alpha) = 0.9739$ is the gamma function and
$z_{\rm c}$ is specified for each sample by the $z_{\rm c} \sim S$
relationship.  Analytic distributions defined in Eq.~\ref{fit_2} are shown in
Fig.~\ref{z_distr} with the solid curves. In all the histograms the model
distribution is astonishingly close to the corresponding best three-parameter
fit represented by the dotted curves. Most deviations visible in some plots are
easily explained by the statistical nature of the problem and/or the large
scale structures present in the catalogs based on the localized sky area
(\citealt{barger02}, \citealt{gilli03}). Systematic shifts between the fits are
present in three histograms below $S = 10^{-14}$ \cgs. It is a direct result of
the assumption of a single $z_{\rm c} = 0.61$ value for all three {\it Chandra}
samples.  It is noticeable that the constant width (in $\log z$) model fits
adequately represent the data over the full range of fluxes.  Small differences
in the width between the three-parameter fits and the final model which are
visible in the NEP, RIXOS and XMS data, apparently do not represent the
systematic effects. In the NEP and RIXOS histograms the final model is slightly
narrower than the individual fits, while in the XMS sample it is wider.

To effectively use the Eq.~\ref{bz} one needs the representation of the source
counts $dN(S)/dS$ over the whole range of fluxes $S$. The parametrization by
\cite{moretti03} adequately suits the present calculations. The smooth
functional form for $N(S)$ proposed by Moretti et al. accurately reproduces the
observed counts below $10^{-11}$ \cgs\ down to {\it Chandra} threshold of
$\sim\!2\times 10^{-17}$ \cgs. Sources within these flux limits generate more
than $90$\,\% of the XRB and smooth extrapolation of the \cite{moretti03}
counts down to $\sim\!3\times 10^{-18}$ \cgs\ is consistent with the entire
XRB.  Substituting all the components into Eq.~\ref{bz} we finally get the
redshift distribution of the XRB photons. It is shown with the solid curve in
the bottom right panel in Fig.~\ref{z_distr}. The same normalization has been
applied to facilitate comparison with the distributions derived for the
individual samples. Points in the plot are discussed below in the
Sec.~\ref{discussion}.

A suitable representation of the $b(z)$ distribution has been found using
a smooth function of the same form as for the individual redshift histograms.
The function:
\begin{equation}
b_{\rm fit}(z) = 5.24\;z^{1.52}\;e^{-z/0.63}\,,
\end{equation}
reproduces the derived distribution of $b(z)$ with the relative error
of less than $4$\,\% for $0.06 < z < 6$. It is shown with dotts
in the bottom right panel (with normalization rescaled to conform to
all the plots in Fig.~\ref{z_distr}).

\section{\large XRB and Supervoids \label{supervoids}}

The distribution of the XRB photons $b(z)$ peaks at redshift $z\approx 1$ and
$50$\,\% of the background originates between the redshifts of $0.4$ and $1.4$
(for $80$\,\% the redshift limits are $0.2$ and $2.1$).  Thus, very large
structures of the matter distribution at redshift within these limits would
generate fluctuations of the integral XRB.  As an example I discuss below the
X-ray signature of the huge void postulated by \cite{rudnick07}.  The arguments
based on the radio survey in favor of the void with a radius of
$\sim\!140$\,Mpc in Eridanus have been questioned (\citealt{smith08}).
Nevertheless, the Integrated Sachs-Wolfe effect operating on extremely large
structures of matter remains a valid explanation of the strongest CMB
fluctuations.

The redshift separation, $\Delta z$, corresponding to the far side and the near
side of the void with diameter $R_o$ centered at redshift $z$ is equal to
 (e.g. \citealt{hogg99}):
\begin{equation}
\Delta z = \frac{2 R_o}{c / H_o} \sqrt{\Omega_m (1+z)^3 + \Omega_\Lambda}\,.
\end{equation}
Using the $b(z)$ distribution we assess the fractional deficit of the XRB
$\delta = \Delta b / b$ created by the completely empty region of size
$280$ Mpc. Such void would generate $|\delta| = 5.7$\,\% and $\delta = 4.9$\,\%
at redshifts $z=0.5$ and $1$, respectively. Assuming spherical shape of the
void, its angular diameter would be $8\fdg 4$ and $4\fdg 7$ at these redshifts.
The XRB depression produced by the void should be compared to the XRB
intrinsic fluctuations
resulting from the discrete nature of sources generating the background.
Assuming purely random distribution of sources, the rms fluctuations
of the XRB are defined by the source counts $N(S)$:
\begin{equation}
\sigma_b = \left[ \int_{S_{\rm min}}^{S_{\rm max}} \;dS\:S^2\:\omega\;
                          \left|\frac{dN(S)}{dS}\right|\;\right]^{\;1/2}\,,
\end{equation}
where $\omega$ is the solid angle subtended by the investigated area. Using the
the \cite{moretti03} counts and $S_{\rm max} = 1\cdot 10^{-11}$ \cgs\ (the
amplitude of the XRB fluctuations is dominated by the contribution of sources
at the bright end of counts), we get $\delta b/b = 0.035$ and $0.024$ for the
circular areas of radius $2^\circ$ and $3^\circ$, respectively. Thus, at
$z=0.5$ the signal-to-noise ratio for the void detection amounts to
$\sim\!2.4$. In the case of $z=1$ the S/N drops to just $1.4$. To reduce the
amplitude of the XRB fluctuations one should remove from the XRB the
contribution of bright sources. If the $S_{\rm max}$ is decreased to $1\cdot
10^{-12}$ \cgs, the significance of the void signal reaches $3.5\,\sigma$ at
$z=0.5$ and $2.0\,\sigma$ at $z=1$. So, only the low redshift voids would
produce the XRB deficits significantly stronger than the statistical
fluctuations.

\section{\large Redshift distribution of the XRB and the AGN evolution \label{agn}}

\begin{figure}
\begin{center}
\includegraphics[width=0.6\textwidth]{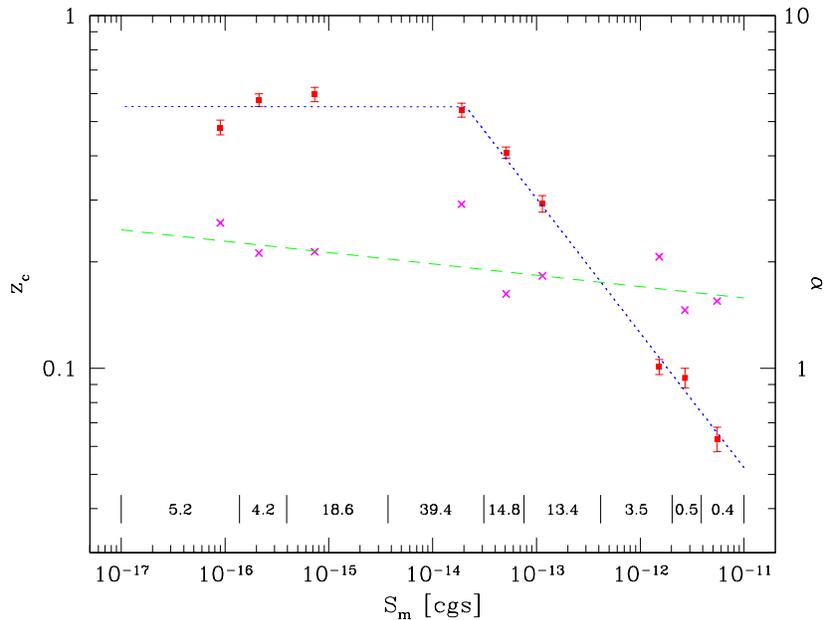}
\caption{\leftskip 10mm \rightskip 10mm
{\small Distributions of best fitted parameters $\alpha$ and $z_{\rm c}$ vs.
the median flux, $S_{\rm m}$, in the AGN samples.  Points with error bars --
$z_{\rm c}$ (left-hand ordinate),  crosses -- $\alpha$ (right-hand ordinate).
The dashed line represents the regression line for the $\alpha \sim S_{\rm m}$
relationship; the $z_{\rm c} \sim S$ function used to model the AGN
contribution to the XRB is shown with the dotted lines (see text for details).
The vertical bars at the bottom mark the effective flux limits of the analyzed
samples; numbers indicate a relative contribution to the XRB generated by the
AGN within these limits.}}
\label{z_c_vs_s_agn}
\end{center}
\end{figure}

To assess the distribution of the XRB generated just by the AGN, I have
repeated all the procedures described in the previous sections using the
samples constructed exclusively from the AGN. The AGN sources are easily
separated from the clusters and nearby normal galaxies. However, the
distinction between the nuclear activity and stellar emitters in the case of
distant and weak sources becomes problematic. Such sources are present in both
{\it Chandra} surveys.  One should keep in mind this limitations in the present
investigation.  Nevertheless, the AGN are a dominating constituent of all the
samples exploited in the paper and even the moderate contamination of the AGN
subsamples with the off-nuclear sources would not affect significantly our
calculations (see below). The numbers in the AGN samples are smaller than in
the full samples and parameter estimates are subject to slightly larger
uncertainties.  Clusters and normal galaxies populate on the average lower
redshift bins and the histograms for the AGN analogous to those in
Fig.~\ref{z_distr} are shifted towards the higher redshifts. We notice also a
weak correlation of the best fit $\alpha$ parameter with the source flux -- a
shape of of the redshift distribution (in $\log z$ bins) varies with $S_{\rm
m}$. In Fig.~\ref{z_c_vs_s_agn} the values of $\alpha$ in the nine samples are
shown with crosses. The regression line of $\log\alpha$ on $\log S$ is used to
fix the value of $\alpha$ for the each sample and to calculate the best fit
parameter $z_{\rm c}$. These new $z_{\rm c}$ are shown in
Fig.~\ref{z_c_vs_s_agn} with the squares. Finally, the best fit line $\log
z_{\rm c} \sim \log S_{\rm m}$ is calculated for six brighter samples to obtain
$z_{\rm c}$ for $S > 1.9\cdot 10^{-15}$ \cgs. A constant $z_{\rm c}$ is assumed
for lower fluxes, and the complete $z_{\rm c} \sim S_{\rm m}$ relationship
adopted for further computations is shown in Fig.~\ref{z_c_vs_s_agn} with the
dotted lines.

The X-ray source counts used in the present case should be limited to the AGN
only. Two other major classes of sources contributing to the counts are
associated with clusters and normal/starburst galaxies. The analytic formula
obtained by \cite{moretti03} quite accurately represents counts of all the
types of extragalactic sources, but the relative contribution of the each class
in the total counts is not well established.  One should notice, however, that
most of the XRB is produced by sources in the middle range of fluxes considered
here, while the cluster contribution is significant only at the bright end of
counts and the normal and starburst galaxies populate mostly the faint end of
counts.  Vertical bars at the bottom of Fig.~\ref{z_c_vs_s_agn} divide
the flux range of $10^{-17} - 10^{-11}$ \cgs\ into $9$ contiguous bands
corresponding approximately to fluxes surveyed by the source samples defined in
the paper.  The numbers between the bars give the relative contribution of each
flux band to the total XRB.

To extract the cluster and starburst galaxies contributions we corrected the
total counts in the following way. In the RBS(b) sample clusters constitute
$40$\,\% of all the extragalactic sources. The slope of the cluster counts at
the bright end amounts approximately to $1.3$ (\citealt{degrandi99}).
Substantially flatter slope than that for the total counts reduces the relative
cluster contribution at lower fluxes. Although the cluster counts are not well
constrained below $\sim\!10^{-12}$ \cgs, their contribution to the total counts
drops at the faint end of counts to a negligible level.  The source counts
attributed to the AGN are assessed by subtracting the cluster counts from the
total counts defined by the \cite{moretti03} formula.  The normal and starburst
galaxies are relatively abundant in the CDFN(f) sample.  Of $241$ extragalactic
sources, $147$ have been classified as `AGN', $51$ as `starburst' and for the
other $43$ the redshift is unknown.  The absolute maximum content of the
non-AGN sources in the CDFN(f) sample amounts to $(51+43)/241 \equiv 39$\,\%,
assuming that all sources with undetermined redshift are starburst galaxies.
The amount of the non-AGN sources at higher flux levels drops quickly. In the
XMS sample none extragalactic source with known redshift has been classified as
normal or starburst galaxy.  The maximum possible contribution of the starburst
galaxies at the low flux levels has been accounted for by flattening the slope
of the \cite{moretti03} counts below $10^{-14}$ \cgs\ to reproduce the
reduction of the AGN in the CDFN(f) sample by $39$\,\%.  The counts modified
this way have been substituted into Eq.~\ref{bz} to obtain the redshift
distribution, $b_{\rm AGN}(z)$.

\begin{figure}
\begin{center}
\includegraphics[width=0.7\textwidth]{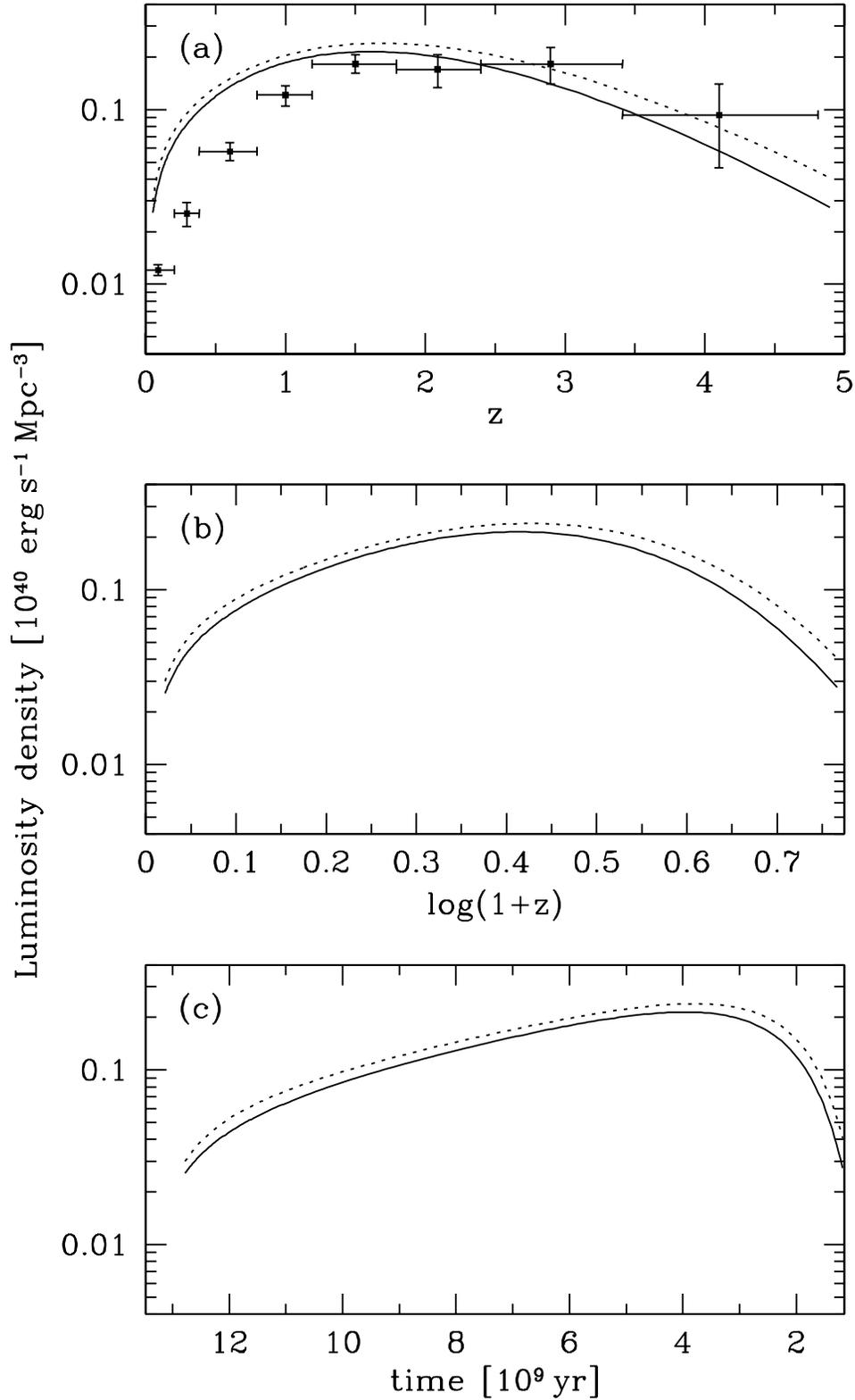}
\caption{\leftskip 10mm \rightskip 10mm
{\small The cosmic evolution of the X-ray luminosity density generated by AGN
as a function of: (a) - redshift $z$, (b) - $\log(1+z)$ and (c) - the cosmic time.
The plots are based on the soft X-ray emission only and do not include
radiation absorbed and re-emitted in different energies. Points with the
error bars in the panel (a) are taken from \cite{hasinger05} and represents
only the type-1 AGN.}}
\label{evolution}
\end{center}
\end{figure}

One can use the observed distribution of the background flux produced by the
AGN to calculate the cosmic history of the luminosity density generated by
these objects.  The $b_{\rm AGN}(z)$ distribution is inserted into
Eq.~\ref{lum_dens} which relates the cosmological evolution of the X-ray
luminosity density, $\varepsilon(z)$, to the redshift distribution of the
background, $b(z)$.  Variations of the luminosity density obtained this way are
shown in Fig.~\ref{evolution} with the solid curves. The data are displayed in
three panels as a function of redshift, $z$, logarithm of $(1+z)$ and the
cosmic time, assuming $t_0 = 13.47\cdot 10^9$ years for the present age of the
Universe\footnote{For the cosmological model defined in Sec.~\ref{intro} and
using the formulae given by \cite{hogg99}.}.

The accuracy of the present $\varepsilon(z)$ estimate depends strongly
on a quality of our $b(z)$ fits. Relatively small numbers of sources
at redshifts below $\sim\!0.03$ and above $\sim\!3$ generate large
statistical fluctuations and weakly constrains the analytic fits $b(z)$
in these redshift ranges. Hence, the present estimates of $\varepsilon(z)$
are also subject to large uncertainties at low and high redshifts.

In order to assess the  importance of the $N(S)$ uncertainties on the present
estimates of $\varepsilon(z)$, I have plotted in Fig.~\ref{lum_dens} with the
dotted curves the $\varepsilon(z)$ function using the original \cite{moretti03}
formula, i.e. assuming no corrections for clusters and starburst galaxies. The
discrepancies between both solutions do not exceed $20$\,\% for redshifts below
$\sim\!3$. It implies that our procedure to isolate the contribution of AGN
from the total counts, albeit crude, does not contribute significantly to the
final errors of $\varepsilon(z)$.

\section{\large Discussion \label{discussion}}

The main objectives of the present investigation, viz. estimates of the redshift
distribution of the XRB photons, $b(z)$, and the evolution of the AGN
luminosity density, $\varepsilon(z)$, have been achieved using the smooth,
analytic fits to the observed source redshift histograms. The present method
is conceptionally simple and computationally straightforward. Unfortunately,
it does not provide error estimates. The major sources of uncertainties
have been indicated in the previous section. Here a quantitative estimate
of the errors is discussed.

The errors of the present measurement of $b(z)$ are generated by the
statistical nature of the investigated material and a chain of approximations
applied to substitute the observed redshift distributions centered on a
selected fluxes by an analytic function $f_S(z)$ continuous in both parameters,
$z$ and $S$.  In fact, the visual inspection of the analytic fits displayed
with the solid curves in Fig.~\ref{z_distr} reveals some deviations from the
redshift histograms.  To estimate the significance of these differences, the
calculations have been performed using the actual histograms shown in
Fig.~\ref{z_distr} with broken solid lines instead of $f_S(z)$. For the each
value of flux $S$ in the range  $10^{-17} - 10^{-11}$ \cgs, the corresponding
redshift distribution has been obtained by the linear interpolation between two
histograms from the samples centered on the median fluxes nearest to $S$.  The
results of this procedure are shown in the lower right panel of
Fig.~\ref{z_distr} with dots.  Generally good agreement between the
distribution of points and the solid curve proves that the analytic
approximations do not introduce perceptible systematic errors in the present
investigation. It appears that the relatively large deviations for three data
points (centered at redshifts: $0.087$, $0.65$, and $1.16$) result purely
from the large scale structures. This is particularly likely for the first
bin ($0.075 < z < 0.1$), where the discrepancy between the fits is
produced entirely by the excess of sources in the localized NEP survey.

One should also notice, that the uncertainties of our main results are only
weakly affected by the limited statistics of the individual samples and
histograms.  This is because the final distributions are obtained by averaging
the individual distributions and this procedure effectively reduces statistical
fluctuations.

In the upper panel of Fig.~\ref{evolution} the AGN emissivity calculated by
\cite{hasinger05} is shown.  The points with the error bars are redrawn here
from their original paper. \cite{hasinger05} apply more stringent criteria to
select sources and use only well defined samples of type-1 AGN. Optically these
objects are identified by the broad Balmer emission lines, while using the
X-ray criteria, they have unabsorbed spectra indicating low intrinsic column
densities. In the present analysis I have included all objects in which the
X-ray emission originates in the active nuclei. Thus, our results cannot be
compared directly to those by \cite{hasinger05}. Nevertheless, despite entirely
different method applied in the present paper, the distributions show good
agreement at redshifts above $\sim\!1$.  Although most of the apparent
discrepancies, which reach a factor of $3$ at $z\approx 0.5$, are probably due
to the distinct selection criteria of both investigations, one cannot exclude
that some differences are caused by unrecognized systematic effects inherent in
one or both methods.

The present method of the luminosity density calculations has also some
disadvantages. In our approach the absolute luminosities of the individual
objects are not determined. Consequently, only the integral luminosity density
is obtained, and the cosmic evolution of any selected AGN luminosity class
has to be studied by means of the standard methods.

\vspace{2mm}
\ni ACKNOWLEDGEMENTS\\
This work has been partially supported by the Polish MNiSW grant N~N203~395934.

\end{document}